\documentclass[a4paper]{jpconf}
\usepackage{graphicx}

\begin{document}
\title{Studies of Scintillator Tiles with SiPM Readout for Imaging Calorimeters}

\author{Frank Simon, for the CALICE Collaboration}

\address{Max-Planck-Institut f\"ur Physik and Excellence Cluster `Universe', Munich, Germany}

\ead{frank.simon@universe-cluster.de}

\begin{abstract}
Imaging hadronic calorimeters with scintillator readout use small scintillator tiles individually read out by silicon photomultipliers to achieve the necessary granularity needed for sophisticated reconstruction algorithms at future collider detectors. For a second generation prototype of the CALICE analog HCAL new, 3 mm thick scintillator tiles with an embedded wavelength shifting fiber and new photon sensor from CPTA are being fabricated. The availability of blue-sensitive SiPMs also allows fiberless coupling of the photon sensor to the tile, a technique requiring modified geometries to achieve a high degree of response uniformity. We discuss results from test bench and from test beam measurements of different scintillator tile geometries as well as prospects for fiberless coupling of photon sensors.

\end{abstract}

\section{Introduction}

The physics program at future high energy $e^+e^-$ colliders imposes stringent requirements on the detector performance. Of particular importance is the precise reconstruction of jets. This is crucial for the measurement of the parameters of particles decaying into hadronic final states, for the separation of $W$ and $Z$ bosons, and for the measurement of missing transverse energy, an important signature for new physics beyond the Standard Model. 

Particle Flow Algorithms \cite{Brient:2002gh, Morgunov:2002pe, Thomson:2009rp} are a promising strategy to achieve an unprecedented jet energy resolution to satisfy the requirements for precision physics at future high energy lepton colliders. These event reconstruction techniques demand extreme granularity in the calorimeters to provide the imaging capabilities necessary to separate showers of individual particles.
In analog hadron sampling calorimeters with scintillator readout this is achieved by using small scintillator cells with typical sizes of 3 x 3 cm$^2$. The first generation prototype analog hadron calorimeter (AHCAL) \cite{Adloff:2010hb} of the CALICE collaboration uses a total of 7608 scintillator tiles, individually read out by silicon photomultipliers (SiPMs) \cite{Bondarenko:2000in}. In each tile a wavelength shifting (WLS) fiber is embedded to match the emission spectrum of the scintillator to the range of optimal efficiency of the photon sensor.

\section{Scintillator Tiles for the CALICE Technical Prototype}
\label{sec:TechPrototype}

\begin{figure}
\centering
\includegraphics[width=0.95\textwidth]{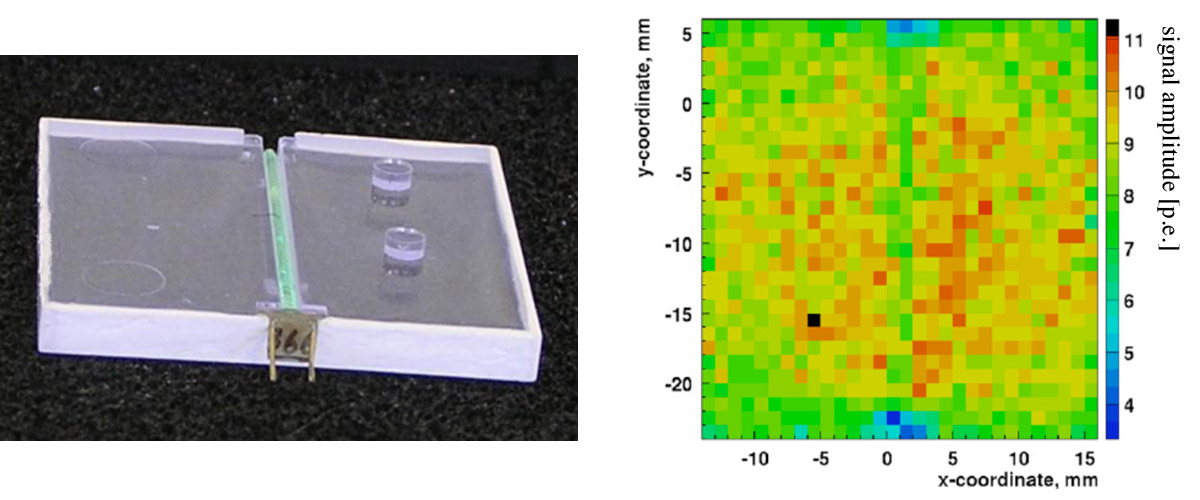}
\caption{{\it Left:} Scintillator tile for the 2$^{nd}$ generation AHCAL prototype, with a thickness of 3 mm and a size of 30 $\times$ 30 mm$^2$, an embedded wavelength shifting fiber, and installed SiPM. {\it Right:} Response map measured with protons.}
\label{fig:TPTile}
\end{figure}

To address technological and integration issues for the hadronic calorimeter at a linear collider detector, a second generation ``technological'' prototype for the analog hadron calorimeter is currently being developed \cite{ErikaCalor}. Key features of this device are fully integrated electronics and a realistic mechanical design. For an increased compactness of the detector, the thickness of the scintillator has been reduced to 3 mm, compared to the 5 mm thickness used in the physics prototype. The size of the tiles, defining the granularity of the detector, is left unchanged at \mbox{30 $\times$ 30 mm$^2$}. The scintillator tiles are fabricated by injection molding. The sides are chemically matted to prevent tile to tile light cross talk. A WLS fiber is embedded in the tile to provide wavelength matching to the SiPM and for improved light collection. The photon sensor is an improved SiPM by CPTA, with 556 pixels arranged in an active area adapted to the cross section of the WLS fiber. For future devices the number of pixels will be increased to 796 pixels for an increased dynamic range. Figure \ref{fig:TPTile} {\it left} shows such a scintillator tile. In addition to the WLS fiber, the photon sensor and the side matting, alignment pins used for the installation of the cells on the printed circuit board housing the front-end electronics are visible.

The new photon sensors and the scintillator tile assemblies have been thoroughly tested with light sources and in particle beams. Figure \ref{fig:TPTile} {\it right} shows a measurement of the response to protons from the ITEP synchrotron as a function of beam position on the scintillator. The SiPM operating point for the calorimeter is chosen to achieve about 10 photo-electrons for a minimum ionizing particle, since this provides a large dynamic range while ensuring an efficiency of around 96\% at a threshold of  0.4 minimum ionizing particle (MIP) equivalent. As the figure shows, a high degree of uniformity is achieved, with a 20\% reduction of the response in the area of the WLS fiber. The positions of the SiPM and of the opposing fiber end result in small areas of significantly reduced sensitivity, as expected.

\section{Laboratory Setup for Uniformity Studies}

\begin{figure}
\centering
\includegraphics[width=0.5\textwidth]{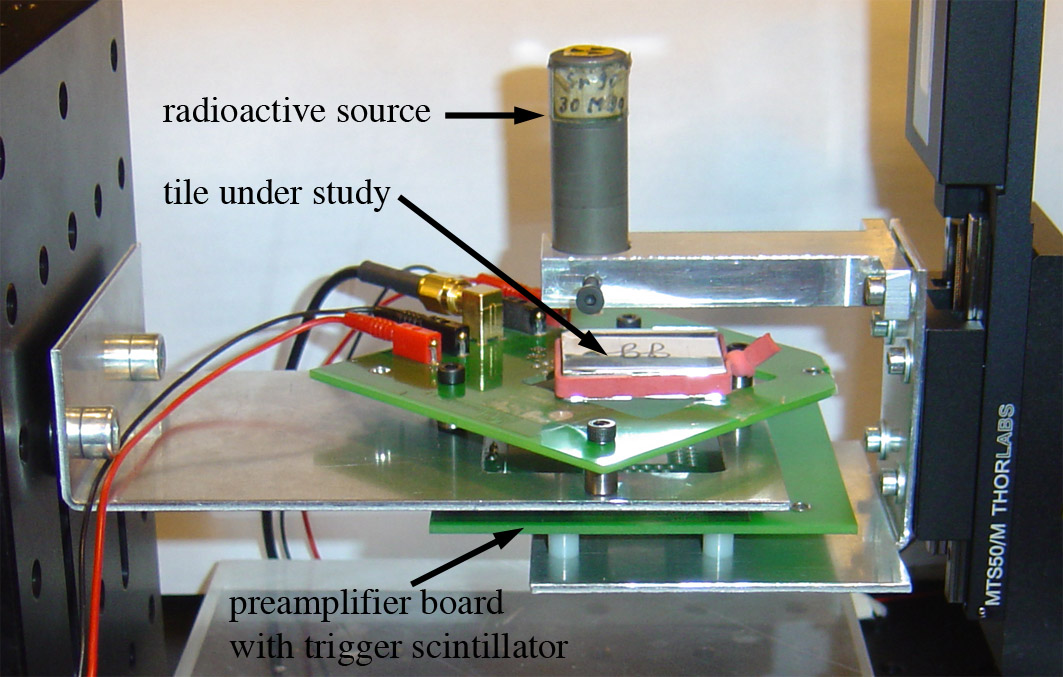} \hfill \includegraphics[width=0.46\textwidth]{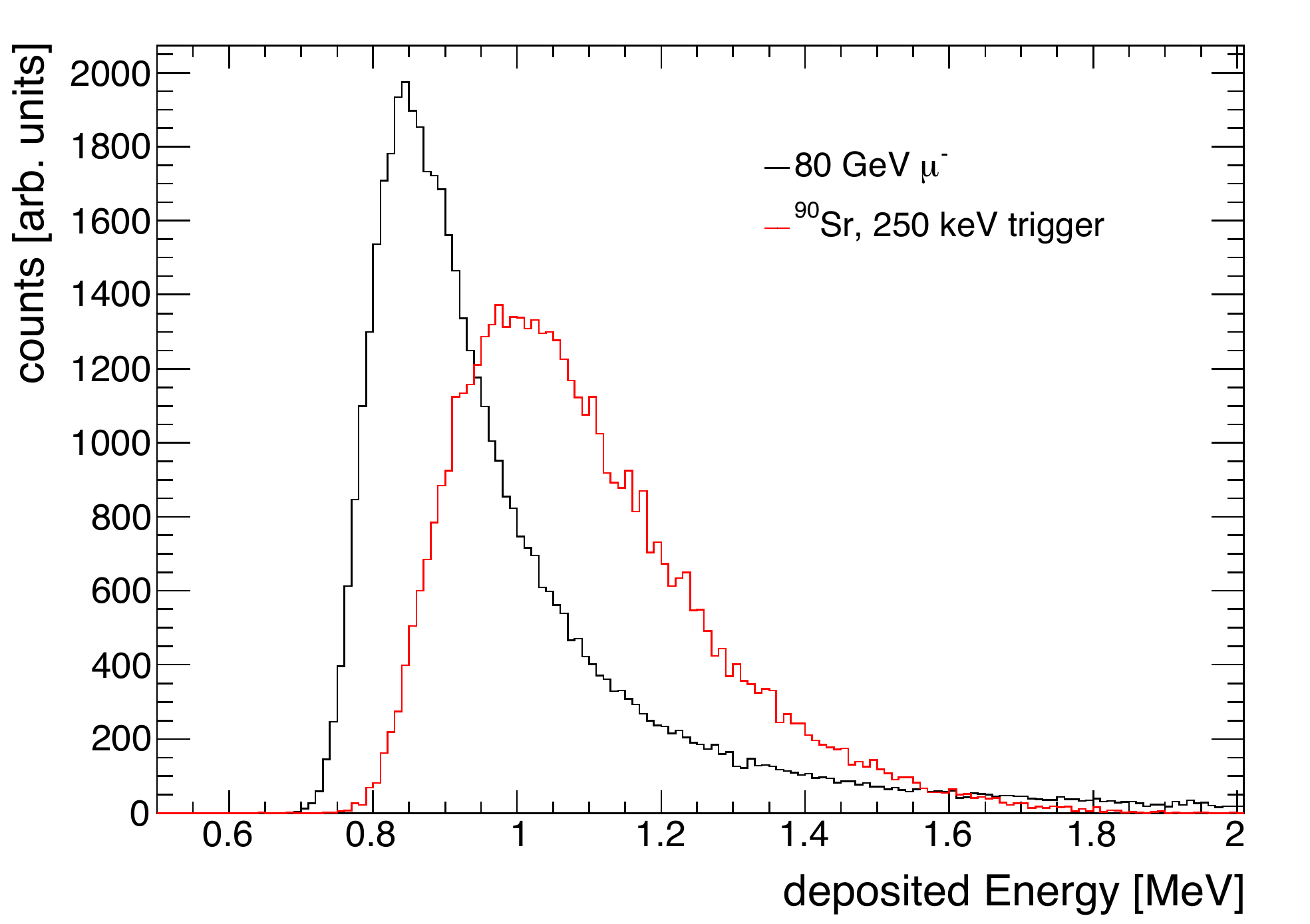}
\caption{{\it Left:} Test setup for detailed response studies of scintillator tiles \cite{Simon:2010hf}.  {\it Right:} Comparison of the energy loss of 80 GeV muons and of penetrating electrons from a $^{90}$Sr source in 5 mm thick scintillator, simulated by {\sc Geant4}. }
\label{fig:TestSetup}
\end{figure}

For detailed studies of the response of small scintillator tiles with SiPM readout, a test setup that scans a $^{90}$Sr source across the scintillator and measures the signal for fully penetrating electrons was constructed \cite{Simon:2010hf}. This setup, shown in Figure \ref{fig:TestSetup} {\it left}, uses high precision stages to position the radioactive source, which is collimated to a spot size of a few mm in diameter. The readout is triggered by a coincidence of the signal from the tile under study and of the signal in a small scintillator cube positioned underneath the tile, which moves together with the source. This ensures that only fully penetrating electrons from the high energy tail of the $^{90}$Sr spectrum, originating from the secondary $^{90}$Y decay, are recorded. The readout is performed with a high-speed oscilloscope, providing precision measurements and single photo-electron resolution.

The energy loss of electrons from the $^{90}$Sr source in the scintillator was determined with a {\sc Geant4} simulation based on a realistic incident energy distribution \cite{BetaSpectra} and a modeling of the trigger by requiring a minimum energy deposit of 250 keV in the trigger scintillator. This was compared to a simulation using 80 GeV muons, as used for calibration in the CALICE test beams at CERN \cite{Adloff:2010hb}. For the muon simulation the absorber plates are included to get a realistic modeling of the energy loss distribution. Figure \ref{fig:TestSetup} {\it right} shows that electrons from the radioactive source deposit approximately 20\% more energy in the scintillator than minimum-ionizing particles and also show a broader distribution with a less pronounced high energy tail. Still, the $^{90}$Sr signals can serve as a good approximation for minimum-ionizing particles, once the higher mean energy deposition is taken into account.

\section{Fiberless Coupling of the Photon Sensors}

With the now readily available blue-sensitive SiPMs, wavelength-matching to the photon sensor with WLS fibers is no longer necessary. This allows a simplification of the construction of the scintillator tiles, and leads to relaxed mechanical tolerances for the coupling of the photon sensor since a precise alignment of the active area of the sensor with the fiber end is not necessary. The elimination of the wavelength shifter also results in a faster response of the scintillator-SiPM system since the additional time constant from the de-excitation in the fiber is no longer present.  However, a fiber embedded in the scintillator serves also as a light collector, leading to a good uniformity of the response over the tile area, which is lost in the case of a simple fiberless coupling of the photon sensor.  

The uniformity of the response over the full active area can be recovered by special shaping of the scintillator \cite{Simon:2010hf, Blazey:2009zz}. In the case of a slit drilled in the side face of the scintillator tile, as discussed in detail in \cite{Simon:2010hf}, the tiles are compatible with the ones used for the 2$^{nd}$ generation prototype discussed in Section \ref{sec:TechPrototype}.  With the scanning setup, 3 mm thick tiles of that type were evaluated, using Hamamatsu Multi-Pixel Photon Counters (MPPCs) with an active area of 1 mm$^2$ and 1600 pixels (type S10362-11-025P), operated at their nominal operation point as given by the manufacturer. Two types of tiles, both with the same geometrical parameters for the slit at the SiPM coupling position, were studied: a tile machined from a plate of BC-420 plastic scintillator and a tile produced with the molding technique used for the CALICE technical prototype tiles. The machined tile was fully enclosed in aluminized reflective foil except for the photon sensor position, while the molded tile had chemically matted sides, with only the top and bottom face covered by reflective foil. In both tiles the slit designed to recover the response uniformity was produced by drilling.

\begin{figure}
\centering
\includegraphics[width=0.95\textwidth]{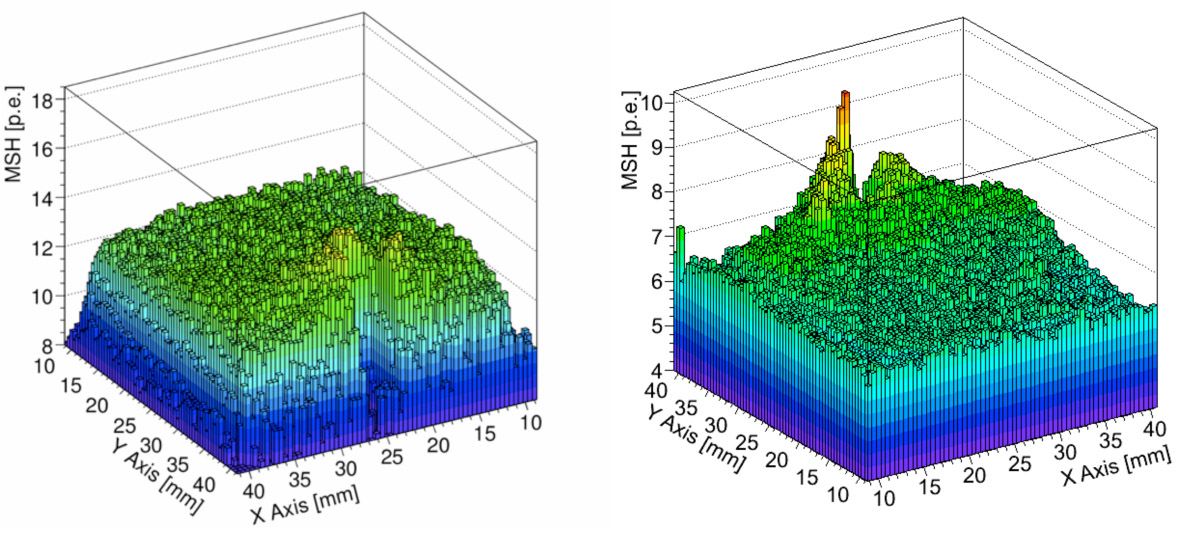}
\caption{Response maps of 3 mm thick scintillator tiles with a coupling slit fiberlessly read out with a Hamamatsu MPPC, showing the mean signal height (MSH). {\it Left:} Machined scintillator fabricated from BC-420, enclosed in reflective foil. {\it Right:} Molded scintillator, with chemically matted sides and reflective foil on top and bottom.}
\label{fig:Uniformity}
\end{figure}

Figure \ref{fig:Uniformity} shows the response map of these two scintillator tiles. The overall mean signal height is drastically different: While the machined tile achieves a mean amplitude of 13 p.e., only 7 p.e. are achieved in the case of the molded tile. It is known that the light yield of the scintillator used for injection molding is about 30\% lower than the one of BC-420, explaining a large part of this difference. In addition, the matting of the sides very likely leads to a decreased light collection efficiency, in particular in areas further away from the photon sensor, further reducing the mean signal height. This loss in signal amplitude can be recovered by using larger photon sensors or SiPMs with fewer, larger pixels and thus higher quantum efficiency, or by increasing the overbias of the sensors. 

The response uniformity of the machined tile is very good, with 82.5\% (94\%) of the active area within 5\% (10\%) of the mean signal height, excluding edge effects. The uniformity of the molded tile is slightly reduced compared to this, with 66\% (87\%) of the active area within 5\% (10\%) of the mean signal height. This reduction is in part due to the reduced reflectivity of the tile sides, which changes the light collection efficiency over the tile surface, and is likely also affected by the low overall signal. The large overshoot at the photon sensor position visible in Figure \ref{fig:Uniformity} is probably due to a coupling problem. Still, an overall satisfactory uniformity was achieved with this first test of fiberless coupling with an injection-molded tile.

\begin{figure}
\centering
\includegraphics[width=0.95\textwidth]{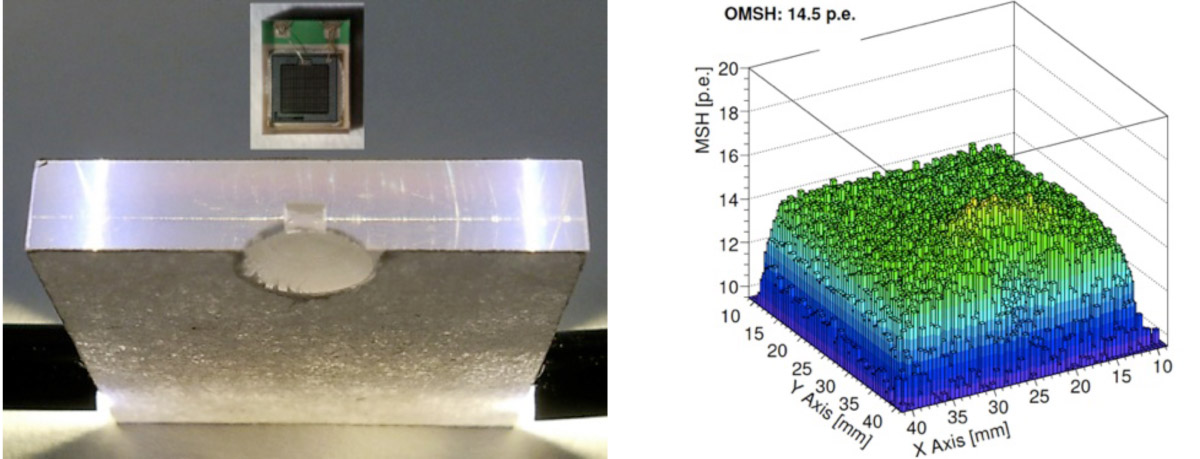}
\caption{{\it Left:} 5 mm thick scintillator tile with a quarter of a spherical hole of 5 mm diameter and a slit for the photon sensor. The photon sensor, a surface-mount Hamamatsu MPPC is shown for reference.{\it Right:} Response map of the scintillator tile, showing the mean signal hight MSH) at each position. The overall mean (OMSH) over the full tile surface is 14.5 photo-electrons.}
\label{fig:UniformityCircular}
\end{figure}

To find tile geometries that provide a high uniformity of the response and are well suited for injection molding further studies have been performed. Figure \ref{fig:UniformityCircular} {\it left} shows a 5 mm thick scintillator tile with a spherical hole with a radius of 5 mm drilled to a depth of 1.8 mm into the bottom face of the tile, with the center of the hole 2 mm from the front edge. In addition to this hole, a slit to accommodate  an MPPC in a compact surface-mount package was machined into the side face. The MPPC, also shown in Figure \ref{fig:UniformityCircular} {\it left}, was installed oriented as shown in the figure, but with the active area facing the tile. Thus, the complete active area of the photon sensor was inside the air volume of the hole in the scintillator. The reduction of the scintillation material at the SiPM position, coupled with diffuse reflection in the drilled hole leads to an excellent uniformity, as shown in Figure \ref{fig:UniformityCircular} {\it right}. It was observed that the performance with a polished hole surface was reduced compared to the rough, diffuse surface obtained by drilling. The overall signal amplitude is reduced by 20\% compared to 5 mm thick tiles with the drilled slit, but an equally good uniformity is achieved with a geometry that is more favorable for injection molding. With the present size of the SiPM package, this design is not directly transferrable to 3 mm thick tiles, but there are no fundamental difficulties to extend this new geometry to thinner scintillators.

\section{Conclusions}

Scintillator tiles with an integrated silicon photomultiplier are key components of highly granular analog hadron calorimeters. From the original design for the CALICE AHCAL physics prototype these systems have been further developed. For the CALICE AHCAL 2$^{nd}$ generation technical prototype, thinner tiles with an embedded wavelength shifting fiber and newly designed SiPMs are being fabricated. These tiles show good response uniformity and the desired signal amplitude. Blue sensitive photon sensors allow the fiberless readout of the scintillator without the addition of a WLS fiber. However, specific modifications of the tile at the coupling position of the photon sensor are necessary to ensure a high degree of uniformity over the full active area. With machined tiles, high uniformity and high signal amplitudes are routinely achieved. First tests with injection molded tiles show encouraging results, suggesting that the mass production of scintillator tiles for imaging hadronic calorimeters based on a fiberless design with directly coupled photon sensors is feasible.  

\section*{References}
\bibliography{CALICE}

\end{document}